%% file: Sadough_Cost_HAL.tex
\documentclass[onecolumn]{IEEEtran}
\usepackage{amsmath,epsfig,amssymb,mathrsfs,psfrag,epsf,enumerate,bm}

\input abrmath.tex

\def\H{{\mathbf H}}
\def\HH{\widehat{{\mathbf H}}}
\def\tg{\mathcal{H}}
\def\y{{\bm y}}
\def\s{{\bm s}}
\def\z{{\bm z}}
\def\W{{ W}}
\def\WT{{\widetilde{ W}}}
\def\esp{{\mathbb E}}
\def\eye{{\mathbb I}}

\def\sig{\mathbf \Sigma}

\def\H{{\mathbf H}}
\def\HH{\widehat{{\mathbf H}}}
\def\tg{\mathcal{H}}
\def\y{{\bm y }}
\def\s{{\bm s }}
\def\z{{\bm z }}
\def\W{{ W}}
\def\WT{{\widetilde{ W}}}
\def\esp{{\mathbb E}}
\def\eye{{\mathbb I}}

\def\sig{\mathbf \Sigma}

\DeclareMathAlphabet{\mathpzc}{OT1}{pzc}{m}{it}

\newcommand{\mb}{\mathbf}

\newcommand{\mc}{\mathcal}

\title{On Optimal Turbo Decoding of Wideband MIMO-OFDM Systems Under Imperfect Channel State Information}
\author{Sajad Sadough$^{\dagger *}$ and Pierre Duhamel$^*$ \vspace{5mm} \\ $^\dagger$Ecole Nationale Sup\'erieure de Techniques Avanc\'ees, 75015 Paris, France\\
$^*$ Laboratoire des Signaux et Syst\`emes, CNRS/Sup\'{e}lec, F-91192 Gif-sur-Yvette, France\\ Email:\{sadough, pierre.duhamel\}@lss.supelec.fr \\[1mm]}
\begin{document}
\maketitle
\begin{abstract}
We consider the decoding of bit interleaved coded modulation (BICM) applied to both multiband and MIMO OFDM systems for typical scenarios where only a noisy (possibly very bad) estimate of the channel is provided by sending a limited number of pilot symbols. First, by using a Bayesian framework involving the channel a {\it posteriori} density, we adopt a practical decoding metric that is robust to the presence of channel estimation errors.
Then this metric is used in the demapping part of BICM multiband and MIMO OFDM receivers. We also compare our results with the performance of a mismatched decoder that replaces the channel by its estimate in the decoding metric.
Numerical results over both realistic UWB and theoretical Rayleigh fading channels show that the proposed method provides significant gain in terms of bit error rate compared to the classical mismatched detector, without introducing any additional complexity. \end{abstract}
\section{Introduction}
\label{sec:intro}
Ultra-Wide-Band (UWB) is defined as any wireless transmission scheme that occupies a bandwidth of more than 25 \% of its center frequency or greater than 500 MHz over the 3.1-10.6 GHz frequency band \cite{fcc}. Multiband Orthogonal Frequency division multiplexing (MB-OFDM) \cite{batra_jour} is a spectrally efficient technique proposed for high data rate, short range UWB applications. This approach uses a conventional OFDM system, combined with bit interleaved coded modulation (BICM) and frequency hopping for improved diversity and multiple access. 

It is well known that multiple-input multiple-output (MIMO) antennas systems is a promising technique for high-speed, spectrally efficient and reliable wireless communications. However, as higher data rates lead to wideband communications, the underlying MIMO channels exhibit strong frequency selectivity. By using orthogonal frequency division multiplexing (OFDM) and applying a proper cyclic prefix (CP), the frequency selective channels are transformed to an equivalent set of frequency-flat subchannels. These considerations motivate the combination of MIMO and OFDM, referred to as MIMO-OFDM, as a promising technology for the future generation of wideband wireless systems \cite{bolskei02}. 

In both MB-OFDM and MIMO-OFDM systems, a typical scenario occurs when the channel is changing so slowly that it is considered time invariant during the transmission of an entire frame.    
In such situations, channel estimation is performed by sending training symbols (pilots) transmitted at the beginning of the information frame while the rest of the frame is decoded based on the estimated channel. Due to the limited number of pilots, the estimate of the channel is imperfect and the receiver has only access to this noisy channel estimate. However, the receiver/decoder metric for any coherent detector, requires knowledge of the exact channel.

A standard sub-optimal technique, known as mismatched ML decoding, consists in replacing the exact channel by its estimate in the receiver metric. Hence, the resulting decoding metric is not adapted to the presence of channel estimation errors (CEE). Although, this scheme is not optimal under imperfect channel estimation, it has been extensively adopted for performance evaluation of single an multi-carrier MIMO systems \cite{garg05}. 
Basically, the transmitter and the receiver strive to construct codes for ensuring reliable communication with a quality of service (QoS), no matter what degree of channel estimation accuracy arises during the transmission. The QoS requirements stand for achieving target rates with small error probability even with very bad channel estimates. 
Thus the presence of CEE, arises the following important question: what type of practical encoder/decoder can achieve the best performance under imperfect channel estimation ? 

As an alternative to the aforementioned mismatch scenario, in \cite{Biglieri_jour} the authors proposed a different decoding metric in the case of space-time decoding of MIMO channels. 
In this paper we see that the metric of \cite{Biglieri_jour} can be derived as the average of the likelihood that would be used if the channel is perfectly known, over all realizations of the channel uncertainty which mitigates the impact of CEE on the decoding performance and provides a robust design. 
The averaging of this metric is performed in the Bayesian framework provided {\it a posteriori} pdf of the perfect channel conditioned on its estimate that characterizes the channel estimation process and matches well the channel knowledge available at the receiver. Based on that metric, we formulate our decoding rule for BICM MB-OFDM and MIMO-OFDM. 
  
The outline of this paper is as follows. In Section \ref{sec:sysmodel} we describe the system model for MB-OFDM and MIMO-OFDM transmission over a frequency selective fading channel. Section \ref{sec:pilot} presents the pilot assisted channel estimation: we specify the statistics of the CEE and then calculate the posterior distribution of the perfect channel conditioned on the estimated channel. This posterior distribution is used in section \ref{sec:metric} to formulate the improved ML decoding metric in the presence of imperfect channel state information at the receiver (CSIR). In section \ref{sec:bicmRX}, we use the general modified metric for soft decoding  BICM MB-OFDM and MIMO-OFDM systems. 
Section \ref{sec:simul} illustrates via simulations the performance of the proposed receiver over both realistic UWB channel environments and uncorrelated Rayleigh fading channels and section \ref{sec:concl} concludes the paper.

Notational conventions are as follows. $\mathbf{I}_N$ represents an ($N \times N$) identity matrix; $\mathbb{E}_{\mathbf{x}}[.]$ refers to expectation with respect to $\mathbf{x}$; $|.|$ and $\|.\|$ and ${\rm Tr}(.)$ denote matrix determinant, Frobenius norm and matrix trace respectively; $(.)^T$ and $(.)^{\mathcal{H}}$ denote vector transpose and Hermitian transpose, respectively.   
\section{TRANSMISSION MODEL}
\label{sec:sysmodel}

\subsection{Multiband OFDM}

\begin{figure}[!t]
\centering
\psfrag{A}{\hspace{-1em}Binary Data}
\psfrag{B}{\hspace{-.5em}{\parbox{6em}{\centering $R=1/2$ Convolutional Encoder}}}
\psfrag{C}{\hspace{-1.2em}{\parbox{6em}{\centering Bit\\ interleaver}}}
\psfrag{D}{\hspace{-1.2em}\parbox{6em}{\centering M-QAM\\ Mapping}}
\psfrag{E}{\hspace{-1.2em}\parbox{6em}{\centering IFFT\\ CP \& GI\\ Addition}}
\psfrag{F}{\hspace{-2em}\parbox{6em}{\centering DAC}}
\psfrag{G}{\hspace{-3em}\parbox{6em}{\centering $\exp\left(j2\pi f_ct\right)$}}
\psfrag{H}{\hspace{-.5em}{TFC: Subband Selection}}
\includegraphics[width=0.6\textwidth,height=5cm]{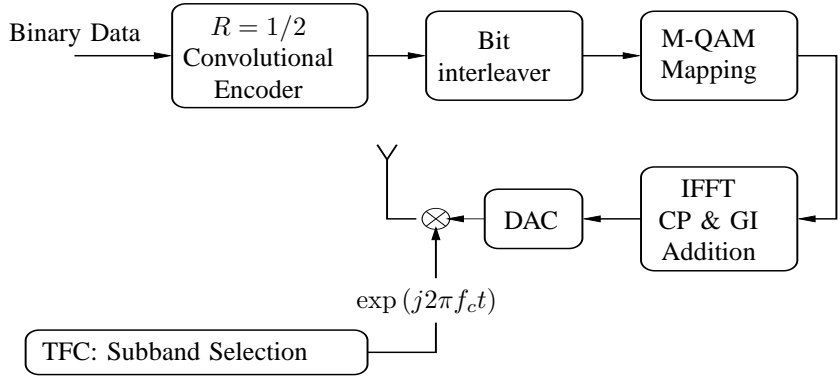}
\caption{TX architecture of the multiband OFDM system.}\label{ofdm_tx}
\end{figure}

A MB-OFDM system divides the spectrum between 3.1 to 10.6 GHz into several
non-overlapping subbands each one occupying approximately 500 MHZ of bandwidth
\cite{batra_jour}. Information is transmitted using OFDM modulation over one of the subbands in a particular time-slot. The transmitter architecture is depicted in Fig. \ref{ofdm_tx}. The binary sequence is convolutionally encoded and then interleaved by a random interleaver. The interleaved bits are gathered in subsequences of $B$ bits $d_k^1,\ldots,d_k^B$ and mapped to complex M-QAM ($M=2^B$) symbols $s_k$ with average energy $E_\mathrm{s}=\mathbb{E}[|s_k|^2]$. At a particular time-slot, a time-frequency code (TFC) selects the center frequency of the subband over which the OFDM symbol is sent. The TFC is used not only to provide frequency diversity but also to distinguish between multiple users.  
We assume an OFDM transmission with $M$ subcarriers, through a frequency selective multipath fading channel, described in discrete-time baseband equivalent form by the taps $\{h_l\}_{l=0}^L$. At the receiver, after removing the cyclic prefix (CP) and performing fast Fourier transform (FFT), a received OFDM symbol over a given subband can be written as
\begin{equation} 
\label{eq:ofdmsys}
        \y = \mathbf{D}_{h_f} \,\s + \z, 
\end{equation}         
where $(M \times 1)$ vectors $\y$ and $\s$ denote received and transmitted symbols, respectively; the noise block $\z$ is assumed to be a circularly symmetric complex Gaussian random vector with distribution $\z \sim \mathcal{CN}(\mathbf{0},{\sigma}^2_z \, \mathbf{I}_M)$; and $\mathbf{D}_{h_f}$ is a diagonal matrix with diagonal elements $\bm{h}_f=[h_f(0),\ldots,h_f(M-1)]^T$, where $h_f(k)=\sum_{l=0}^{L} h_l e^{-j 2 \pi k l / M}$.  

\subsection{MIMO-OFDM}
\begin{figure}[!t]
\centering
\includegraphics[width=0.6 \textwidth,height=0.2\textheight]{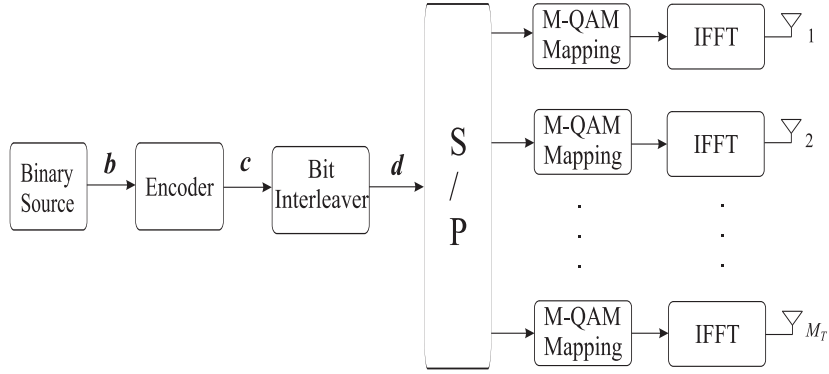}
\caption{Block diagram of MIMO-OFDM transmission scheme.}\label{mimofdm_tx}
\end{figure}

We consider a single-user MIMO-OFDM communication system over a memoryless frequency selective Rayleigh fading channel. The system consists of $M_T$ transmit and $M_R$ receive antennas ($M_R \geq M_T$), and $M$ is the total number of subcarriers. Fig. \ref{mimofdm_tx} depicts the BICM coding scheme used at the transmitter. The binary data sequence ${\bm b}$ are encoded by a non-recursive non-systematic convolutional (NRNSC) code before being interleaved by a quasi-random interleaver. The output bits ${\bm d}$ are multiplexed to $M_T$ substreams and mapped to complex $\widetilde{\rm M}$-QAM symbols before being modulated by the OFDM modulator and transmitted through $M_T$ antennas.

Let $\s$ be the $M M_T\times 1$ vector containing the OFDM symbols transmitted simultaneously over $M_T$ antennas. The symbols are assumed to be independent identically distributed (i.i.d.) with zero mean and unit covariance matrix $\sig_s=\esp[\s\s^\tg]=\eye_{M\times{M_T}}$. Assuming an invariant channel over a frame of $L$ symbols, the received vector $\y$ at a given time index $l$ (omitted for brevity) can be written as
\begin{equation}
\label{eq:model1}
\y = \H \, \s + \z
\end{equation}
where $\H$ is a $M M_R \times M M_T$ block diagonal channel matrix containing the frequency response of the MIMO channels and the noise vector $\z$ is assumed to be a zero-mean circularly symmetric complex Gaussian  (ZMCSCG) random vector with covariance matrix $\sig_z \triangleq \mathbb{E}(\z \z^{\mathcal{H}})=\sigma^2_z \mathbb{I}_{M\times{M_R}}$. We assume that for each frame, a different realization of $\H$ independent of both $\s$ and $\z$ is drawn and remains constant during this frame.
The MIMO-OFDM channel can be decoupled into $M$ frequency flat MIMO channels by exploiting the block diagonal structure in \eqref{eq:model1} which can be rewritten as a set of $M$ equations that contains only one subcarrier each
\begin{equation}
\label{eq:model2}
\y_k = \H_k \, \s_k + \z_k \;\;\;\; k=1,...,M,
\end{equation}
where $\H = {\rm diag}{[\H_1 \H_2 \,...\, \H_M]}$, $\y^T=[\y_1^T \, ...\, \y_M^T]$, $\s^T=[\s_1^T \, ...\, \s_M^T]$ and $\z^T=[\z_1^T \, ...\, \z_M^T]$. The architecture of \eqref{eq:model2} constitutes the basis for the study in this paper.

Hereafter, without loss of generality, we adopt to the more general MIMO-OFDM model of equation \eqref{eq:model1}, where the MB-OFDM case can be deduced by setting $M_T=M_R=1$.
\vspace{3mm}
\\
\section{Pilot Based Channel Estimation}
\label{sec:pilot}
In practical situations, the receiver has only access to a noisy estimate of the channel that differs from the true channel. Under the assumption of a time-invariant channel over the entire transmitted frame, channel estimation is usually performed on the basis of known training (pilot) symbols, transmitted at the beginning of each frame.

We consider the estimation of channel matrix $\H_k$ via the transmission of $N$ training vectors $\s_{_{T,i}}$,   ($i=1,...,N$).
According to \eqref{eq:model1}, when pilot symbols are transmitted we receive
 \begin{equation}
   \label{eq:model3}
         \mathbf{Y}_T = \H_k \,\mathbf{S}_T + \mathbf{Z}_T
  \end{equation}
where each column of the $M_T \times N$ matrix $\mathbf{S}_T=[\s_{_{T,1}}|...|\s_{_{T,N}}]$ contains one pilot symbol and the noise $\mathbf{Z}_T$ has the same distribution as the noise $\z_k$. The average energy of the training symbols is $P_T = \frac{1}{N M_T}{\rm Tr}\big(\mb{S}_T \mb{S}_T^\tg\big)$.
The ML estimate of $\H_k$ is obtained by minimizing $\|\mathbf{Y}_T-\H_k \,\mathbf{S}_T\|^2$ with respect to $\H_k$. We have
\begin{equation}
\label{eq:model4}
\HH_k^{\rm ML}=\mathbf{Y}_T \,\mathbf{S}_T^\tg\,(\mathbf{S}_T\mathbf{S}_T^\tg)^{-1}=\H_k + \mathbf{\mathcal{E}}
\end{equation}
where $\mathbf{\mathcal{E}}= \mathbf{Z}_T \mathbf{S}_T^\tg\,(\mathbf{S}_T\mathbf{S}_T^\tg)^{-1}$ denotes the estimation error matrix. When the training sequence is orthogonal ($\mathbf{S}_T \mathbf{S}_T^\tg = N P_T \eye_{M_T}$), the $j$-th row $\mathbf{\mathcal{E}}
_j$ of the estimation error matrix reduces to a white noise vector with covariance matrix $\sig_{\mathcal{E},j}= \esp\big[\boldsymbol{\mathcal{E}}_j^\tg \boldsymbol{\mathcal{E}}_j\big]=\sigma_{\mathcal{E},k}^2 \eye_{M_T}$, where $\sigma_{\mathcal{E},k}^2=SNR_T^{-1}\triangleq \frac{N P_T}{\sigma^2_z}$.
By assuming that the channel matrix $\H_k$ has the {\it prior} distribution $\H_k \sim \mathcal{CN}(\mathbf{0},\eye_{M_T} \otimes \sig_{H,k})$ and choosing an orthogonal training sequence, we can derive the posterior distribution of the perfect channel conditioned on the estimated channel as
\begin{equation}
\label{eq:model5}
f(\H_k|\HH_k^{\rm ML}) = \mathcal{CN}( \sig_{\Delta} \HH_{\rm ML} ,\, \eye_{M_T} \otimes \sig_{\Delta} \sig_{ \mathcal{E} } )
\end{equation}
where $\sig_{\Delta}=\sig_{H,k}(\sig_{\mathcal{E}}+\sig_{H,k})^{-1}=\delta \eye_{M_R}$ and $\delta=\frac{SNR_T \sigma^2_h}{(SNR_T \sigma^2_h+1)}$.

The availability of the estimation error distribution constitutes an interesting feature of pilot assisted channel estimation that we used to derive the posterior distribution \eqref{eq:model5}. This distribution is exploited in the next section, in the formulation of a modified metric for improving the detection performance under imperfect channel estimation. 
\vspace{3mm}
\section{Maximum-Likelihood Detection in the Presence of Channel Estimation Errors}
\label{sec:metric}
\subsection{Mismatched ML Detection}
\label{subsec:mis}
It is well known that under i.i.d. Gaussian noise, detecting $\s_k$ is given by maximizing the likelihood function $\W(\y_k|\s_k,\H_k)$ which is equivalent to minimizing the Euclidean distance $\mathcal{D}_{\rm ML}$
\begin{equation}
\label{eq:metric1}
\hat{\s}_k^{\rm ML}(\H_k)= \argmin_{\s_k \in \, \mathbb{C}^{M_T \times 1}}\, \big \{ \,\mathcal{D}_{\rm ML}(\s_k,\y_k,\H_k) \big \},
\end{equation}
where
$\mathcal{D}_{\rm ML}(\s_k,\y_k,\H_k) \triangleq - \ln \W(\y_k|\s_k,\H_k)\propto \|\y_k - \H_k \s_k \|^2$.
Since the above detection rule requires the knowledge of the {\it perfect} channel matrix $\H_k$, one sub-optimal approach, referred to as mismatched detection, consists in replacing the exact channel by its estimate in the receiver metric as
\begin{equation}
\label{eq:metric3}
\hat{\s}_k^{\rm ML}(\HH_k)= \argmin_{\s_k \in \, \mathbb{C}^{M_T \times 1}}\, \big \{ \,\|\y_k - \H_k \s_k \|^2 \big \}_{\big |_{\H_k=\HH_k} }
\end{equation}

\subsection{Improved ML Detection for imperfect CSIR}
\label{subsec:mod}
The Bayesian framework introduced previously let us to define a new likelihood function $\WT(\y_k|\HH_k,\s_k)$ by averaging the likelihood that would be used if the channel were perfectly known ($W(\y_k|H_k,\s_k)$), over all realizations of the perfect channel for a given estimated channel state (the posterior distribution \eqref{eq:model5}). This yields
\begin{align}
	\label{eq:metric4}
             \WT(\y_k|\HH_k,\s_k) &= \esp_{\H_k | \HH_k} \big[ \W(\y_k|\H_k,\s_k) \big | \, \HH_k \big] \notag \\
             &= \int_{\H_k \in \mathbb{C}^{M_R \times M_T}} \W(\y_k|\H_k,\s_k) \; f(\H_k|\HH_k) \;\;\; {\rm d}\H_k
\end{align}
Since both $\W(\y_k|\H_k,\s_k)$ and $f(\H_k|\HH_k)$ are Gaussian densities, it is easy to show that
$\WT(\y_k|\HH_k,\s_k) = \mathcal{CN}(\boldsymbol{\mu}_{_\mathcal{M}},\sig_{_\mathcal{M}})$                              with \cite{sadough06}
\begin{equation}
  \label{eq:metric5}
  \left\{\begin{array}{ll}
       \boldsymbol{\mu}_{_\mathcal{M}} &= \delta \, \HH_k \, \s_k, \\
       \sig_{_\mathcal{M}} &= \sig_z  + \delta \, \sig_{\mathcal{E}} \, \| \s_k \|^2.
    \end{array} \right.
\end{equation}
Now the ML estimate of $\s_k$ can be formulated as
\begin{equation}
\label{eq:metric6}
  \hat{\s}_k^{\mathcal{M}}(\HH_k)= \argmin_{\s_k \in \, \mathbb{C}^{M_T \times 1}}\, \big \{ \,\mathcal{D}_{\mathcal{M}}(\s_k,\y_k,\H_k) \big \},
\end{equation}
with
\begin{align}
\label{eq:metric7}
& \mathcal{D}_{_\mathcal{M}}(\s_k,\y_k,\HH_k) \triangleq - \ln \WT(\y_k|\s_k,\HH_k) \notag \\
&= M_R \ln \, \pi (\sigma^2_z + \delta \, \sigma^2_\mathcal{E}\, \|\s_k\|^2)+ \frac{\|\y_k-\delta\,\HH_k\,\s_k\|^2}{\sigma^2_z+\delta\,\sigma^2_\mathcal{E}\,\|\s_k\|^2}
\end{align}
being the new ML decision metric under CEE.

We note that when exact channel is available ($\HH_k=\H_k$), the posterior expectation of \eqref{eq:metric4} becomes equivalent to replacing $\H_k$ by $\HH_k$ in $\W(\y_k|\H_k,\s_k)$ and consequently the the two metrics $\mathcal{D}_{_\mathcal{M}}$ and $\mathcal{D}_{\rm ML}$ coincides. Under near perfect CSIR, occurred either when $\sigma^2_{\mathcal{E}} \to 0$ or when $N \to \infty$, we have $\delta \to 1$, $\delta \sigma^2_{\mathcal{E}} \to 0$,
$\boldsymbol{\mu}_{_\mathcal{M}} \to \HH_k \, \s$ and $\sig_{_\mathcal{M}} \to \sigma^2_z \eye_{M_R}$.
Consequently, the improved metric $\mathcal{D}_{_\mathcal{M}}$ behaves similarly to the classical Euclidean distance metric $\mathcal{D}_{\rm ML}$.
\begin{figure}[!t]
\centering
\includegraphics[width=0.6\textwidth,height=0.33\textheight]{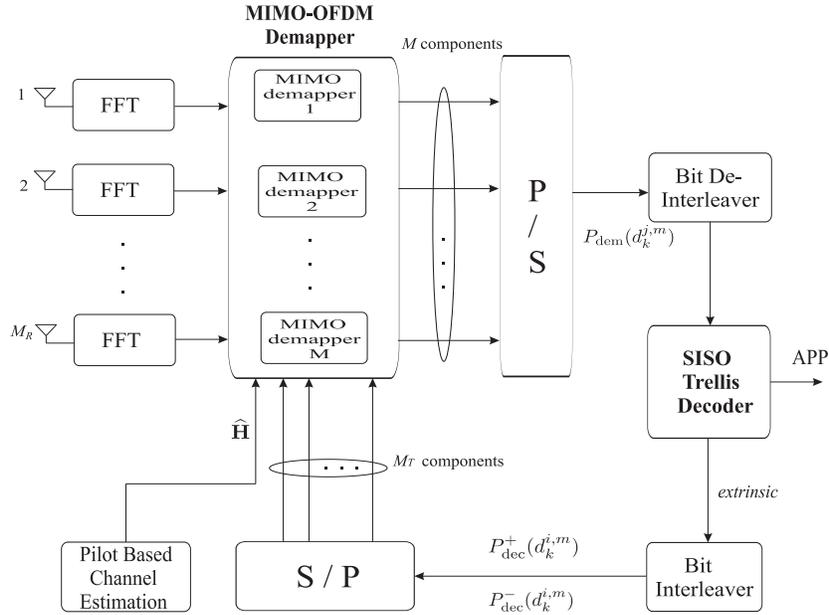}
\caption{Block diagram of MIMO-OFDM BICM receiver.}\label{mimofdm_rx}
\end{figure}
\vspace{3mm}
\section{Iterative Decoding of BICM MIMO-OFDM Based On Imperfect CSIR}
\label{sec:bicmRX}

As a practical application of the general decoding metric \eqref{eq:metric7} proposed in \cite{Biglieri_jour}, we consider soft iterative decoding of BICM MIMO-OFDM under imperfect CSIR. This problem has been addressed in \cite{boutros00} under the assumption of perfect CSIR. Here, without going into the details, we extend the results of \cite{sadough06} to MIMO-OFDM block fading channels estimated by a finite number of training symbols.

As shown in Fig. \ref{mimofdm_rx}, the BICM receiver consists of a bunch of demodulator/demapper, a de-interleaver
and a soft-input-soft-output (SISO) decoder. Let $d_k^{j,m}$ be the $m$-th coded and interleaved bit ($m=1,2,...,\log_2 \widetilde{\rm M}$) of the constellation symbol $\s_k$ at the the $i$-th transmit antenna and the $k$-th subcarrier.
We denote by $L(d_k^{j,m})$ the coded log-likelihood ratio (LLR) value of the bit $d_k^{j,m}$.
At each decoding iteration, the LLR values conditioned on the CSIR are given by
\begin{equation}
\label{eq:mimoRx1}
L(d_k^{j,m})=\log \frac{P_{\rm dem}(d_k^{j,m}= 1)|\y_k,\H_k)}{P_{\rm dem}(d_k^{j,m}=0|\y_k,\H_k)}.
\end{equation}
We have (see \cite{sadough06} and references therein)
\begin{equation}
\label{eq:mimoRx2}
L(d_k^{j,m})= \log \frac{\sum \limits_{\s_k:\, d_k^{j,m}=1} \exp\big\{-\mc{D}_{{\rm ML}}(\s_k,\y_k,\H_k) \big\} \prod \limits_{\stackrel{i=1}{i\neq j}}^B P^+_{\rm dec}\big(d_k^{i,m}\big)}{\sum \limits_{\s_k:\, d_k^{j,m}=0} \exp \big\{-\mc{D}_{{\rm ML}}(\s_k,\y_k,\H_k) \big\} \prod \limits_{\stackrel{i=1}{i\neq j}}^B P^-_{\rm dec}\big(d_k^{i,m}\big)}
\end{equation}
where $P^+_{\rm dec}(d_k^{i,m})$ and $P^-_{\rm dec}(d_k^{i,m})$ are {\it extrinsic} information coming from the SISO decoder.

Notice that the metric $\mc{D}_{\rm ML}(\s_k,\y_k,\H_k)$ involved in \eqref{eq:mimoRx2} requires the knowledge of the perfect channel $\H_k$ of which the receiver has {\it solely} an estimate $\HH_k$.
Contrary to the mismatch approach that replaces the perfect channel by its estimate, we propose to use the improved decoding metric $\mc{D}_{_{\mc M}}(\s_k,\y_k,\HH_k)$ in \eqref{eq:mimoRx2}, in order to derive a new demaping rule adapted to the imperfect channel available at the receiver.
The decoder accepts the LLRs
of all coded bits and employs the well known forward-backward algorithm \cite{bcjr} to compute the LLRs of information bits, which are used for the decision.
\vspace{3mm}
\section{Simulation results}
\label{sec:simul}
Simulations have been carried out in the context of IEEE802.15 wireless PAN \cite{norme_mb}: $M=100$ data subcarriers along with 32 CP samples compose one OFDM symbol. 
Information bits are encoded by a rate $R=1/2$ convolutional encoder with constraint length 3 defined in octal form by (5,7). The interleaver is a pseudo-random one operating over the entire frame and the mapping is 16-QAM. 
Throughout the simulations, each frame is assumed to consists of one OFDM symbol with 100 subcarriers belonging to a 16-QAM constellation with Gray or set partion (SP) labeling. The interleaver is a pseudo-random one operating over the entire frame with size $M\! \cdot\! M_T\! \cdot\! \log_2(\widetilde{M})$ bits.
For each transmitted frame, a different realization of the realistic UWB channel model specified in \cite{chanReport} or Rayleigh distributed channel has been drawn and remains constant during the whole frame. Besides, it is assumed that the average pilot symbol energy is equal to the average data symbol energy. Moreover, the number of decoding iterations are set to either 3 or 4. 

\begin{figure}[!htb] 
\centering
\includegraphics[width=0.7\textwidth,height=0.4\textheight]{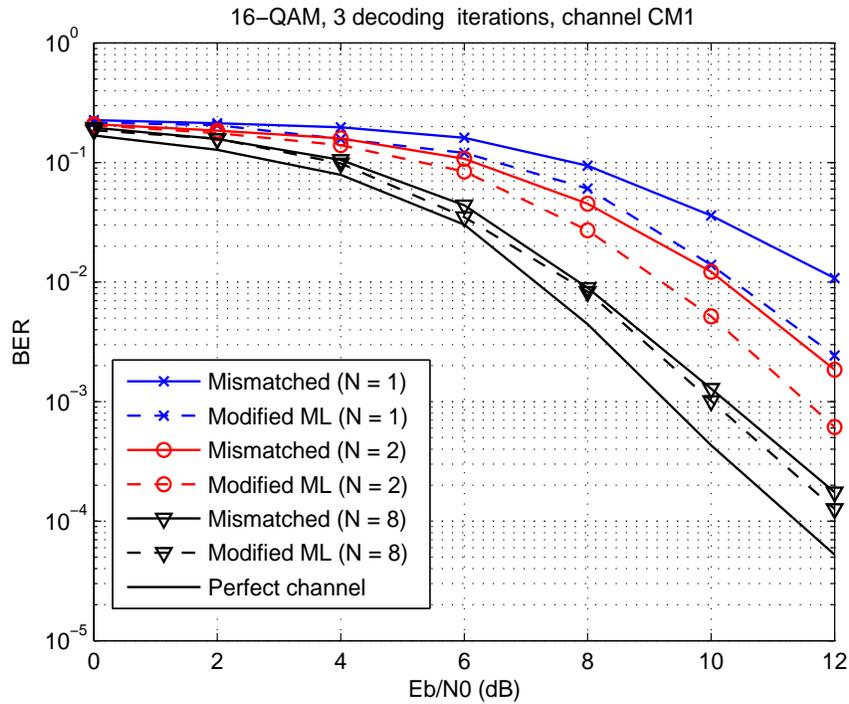} 
\caption{BER performance of the proposed decoder over the LOS UWB channel CM1 for various training sequence lengths, 16-QAM with Gray labeling.}\label{fig3}
\end{figure}

\begin{figure}[!htb] 
\centering
\includegraphics[width=0.7\textwidth,height=0.4\textheight]{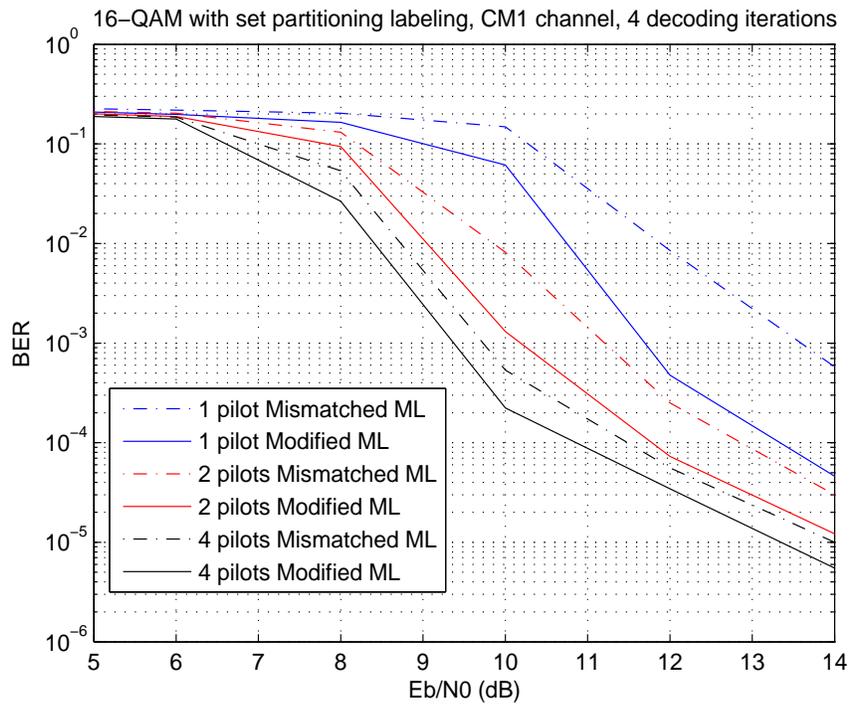} 
\caption{BER performance of the proposed decoder over the LOS UWB channel CM1 for various training sequence lengths, 16-QAM with Set Partition labeling.}\label{ber_SP}
\end{figure}

\begin{figure}[!htb] 
\centering
\includegraphics[width=0.7\textwidth,height=0.4\textheight]{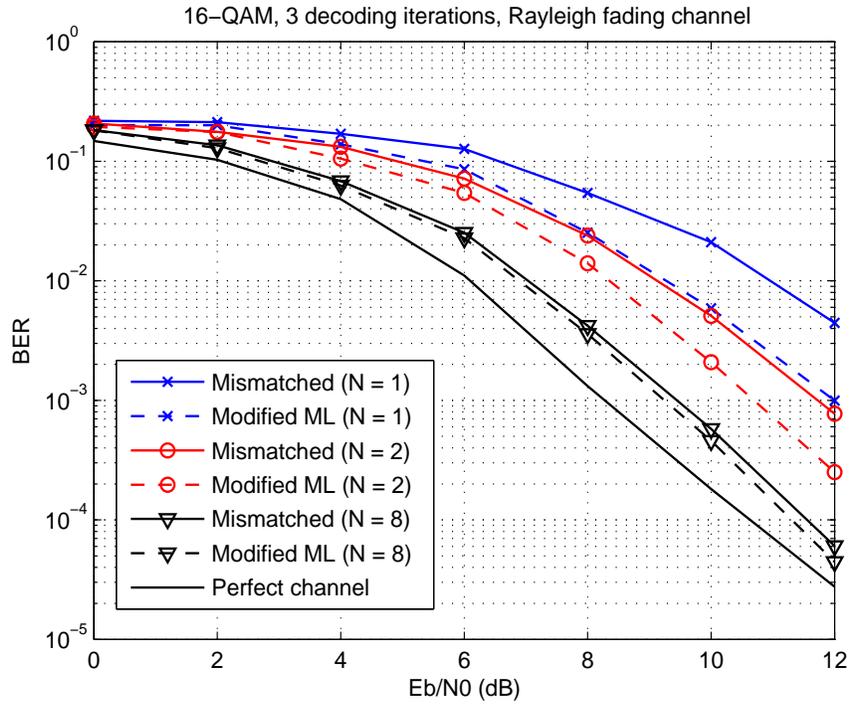} 
\caption{BER performance of the proposed decoder over Rayleigh fading channel for various training sequence lengths, 16-QAM with Gray labeling.}\label{fig4}
\end{figure}

Figure \ref{fig3} depicts the bit error rate (BER) performance gain that is obtained by decoding BICM MB-OFDM with the modified ML decoder in the context of LOS CM1 channel, where the channel estimation is performed by sending different training sequences with lengths $N=\{1, 2, 8\}$ per frame. Similar plots are shown in figure \ref{ber_SP} with SP labeling. 
It can be noticed that for both Gray and SP labeling, the proposed decoder outperforms the mismatched decoder especially when few numbers of pilot symbols are dedicated for channel estimation. Note that the modified decoder with $N=1$ pilot performs very close to the mismatched decoder with $N=2$ pilots. 
For comparison, results obtained with theoretical Rayleigh fading channel are illustrated in figure \ref{fig4}. It can be observed that for $N=2$, the SNR to obtain a BER of $10^{-3}$ is reduced by about $1.5$ dB if the modified ML decoding is used instead of the mismatched approach.     

\begin{figure}[!htb] 
\centering
\includegraphics[width=0.7\textwidth,height=0.4\textheight]{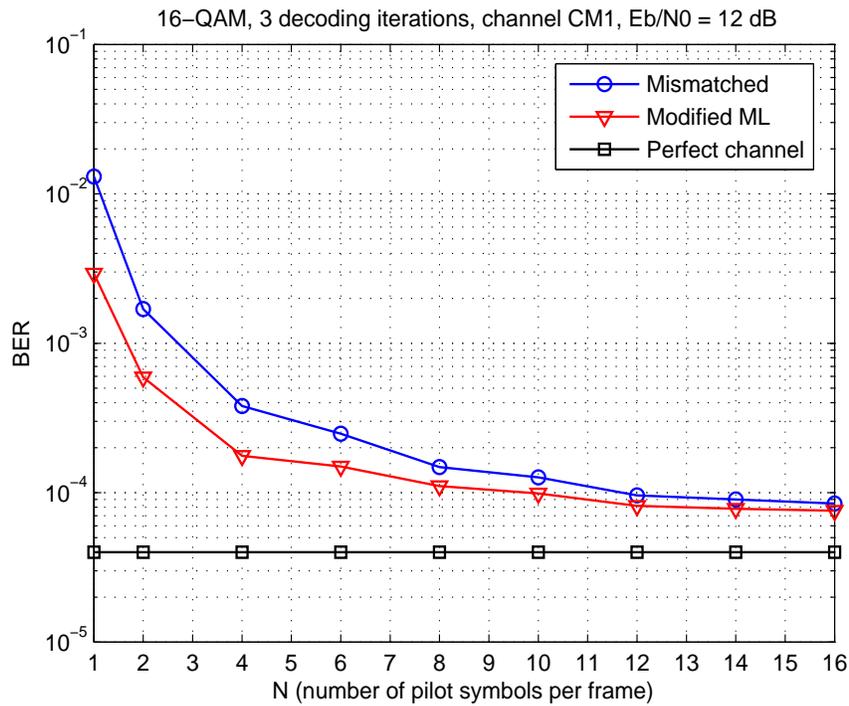} 
\caption{BER versus the number of pilot symbols at $E_b/N_0=12$ dB over the LOS channel CM1.}\label{fig5}
\end{figure}

Figure \ref{fig5} shows the BER performance versus the training sequence length $N$ at a fixed $Eb/N_0$ of $12$ dB for the CM1 channel with 16-QAM and Gray labeling. This allows to evaluate the number of training sequence necessary to achieve a certain BER. We observe that at $Eb/N_0=12$ dB, the modified ML decoder requires $N=9$ pilot symbols per frame to achieve a BER of $10^{-4}$ while the mismatched decoder attains this BER for $N=12$ pilot symbols. Besides, this has been outlined in Section IV, for large training sequence lengths ($N \geq 12$), both decoders have close performance.   

\begin{figure}[!htb]
\centering
\includegraphics[width=0.7\textwidth,height=0.4\textheight]{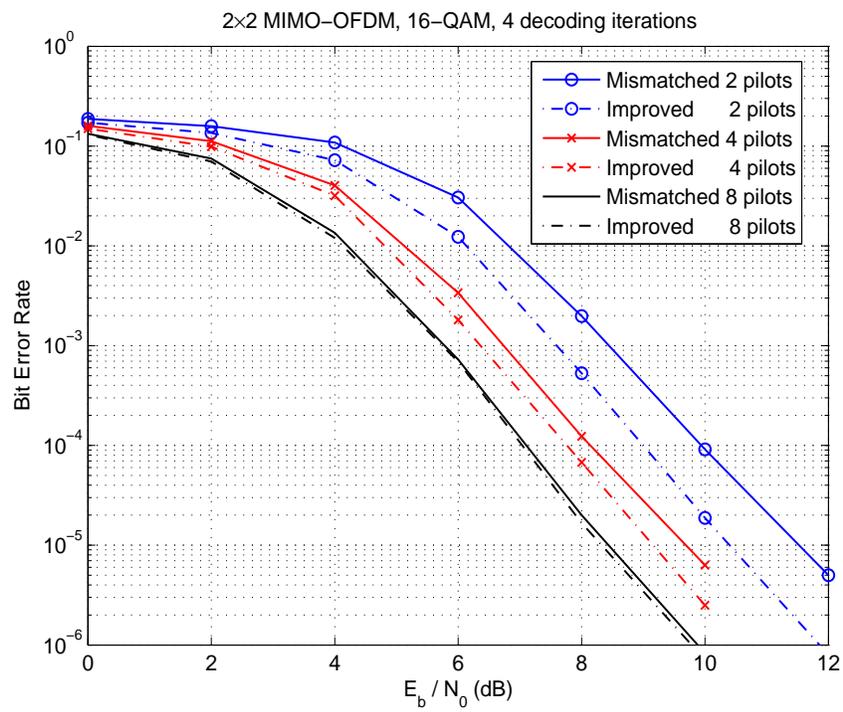}
\caption{BER performance of the proposed and mismatched decoders in the case of a $2\times2$ MIMO-OFDM Rayleigh fading channel with $M=100$ subcarriers for training sequence length $N\in\{2, 4, 8\}$.} \label{bermimofdm}
\end{figure}

Figure \ref{bermimofdm} depicts the BER performance over a $2\times2$ MIMO-OFDM channel estimated by $N\in\{2,4,8\}$ pilot symbols per frame. As observed, the increase in the required $E_b/N_0$ caused by CEE is an
important effect of imperfect CSIR in the case of mismatched ML decoding.
The figure shows that the SNR to obtain a BER of $10^{-5}$ with $N=2$ pilots is reduced by about $1.5$ dB if the improved decoder is used instead of the mismatched decoder. We also notice that the performance loss of the mismatched receiver with respect to the derived receiver becomes insignificant for $N\geq8$. This can be explained from the expression of the metric \eqref{eq:metric7}, where we note that by increasing the number of pilot symbols, this expression tends to the classical Euclidean distance metric. However, the proposed decoder outperforms the mismatched decoder especially when few numbers of pilot symbols are dedicated for channel estimation.
\vspace{3mm}
\section{Conclusion}
\label{sec:concl}
This paper studied the problem of ML reception when the receiver has only access to a noisy estimate of the channel in the case of pilot assisted channel estimation. 
By using the statistics of the estimation error, we adopted a modified ML criterion that is expressed in terms of the estimated channel coefficient.  
This modified metric let us to formulate appropriate branch metrics for decoding BICM MB-OFDM with imperfect channel knowledge.
Simulation results conducted over realistic UWB channels, indicate that mismatched decoding is quite sub-optimal for short training sequence and confirmed the adequacy of the adopted decoding rule in the presence of channel estimation errors. This was obtained without introducing any additional complexity.    

\section*{Acknowledgment}
The authors would like to acknowledge discussions with Pablo Piantanida.

\bibliographystyle{IEEEbib}
\bibliography{bibli-cost}
\vspace{10mm}

\end{document}

%% file: abrmath.tex
\def\esp{{\mathrm{E}}\,}

\newsavebox{\fminibox}
\newlength{\fminilength}

  \def\+{^\dagger}

\def\W{,\thickspace}

\def\nequiv{\not\kern-.05em\equiv}
\def\egal{\kern-.5em=\kern-.5em}        
\def\propt{\kern-.2em\propto\kern-.2em} 

\def\argmin{\mathop{\mathrm{arg\,min}}} 
\def\intdouble{\int\kern-0.3em\int}
\def\inttriple{\int\kern-0.3em\int\kern-0.3em\int}

\def\rond#1{\overset{\kern-0.33em~_\circ}{#1}}
\def\rondit[#1]#2{\overset{\kern#1~_\circ}{#2}}


%% file: Sadough_Cost_HAL.bbl
\begin{thebibliography}{10}

\bibitem{fcc}
FCC,
\newblock ``First report and order, revision of part 15 of the commission's
  rules regarding ultra-wideband transmission systems,''
\newblock Tech. {R}ep., Feb. 2004.

\bibitem{batra_jour}
A.~Batra, J.~Balakrishnan, G.~R. Aiello, J.~R. Foerster, and A.~Dabak,
\newblock ``Design of multiband {OFDM} system for realistic {UWB} channel
  environments,''
\newblock {\em IEEE Transaction on Microwave Theory and Techniques}, vol. 52,
  pp. 2123--2138, september 2004.

\bibitem{bolskei02}
H.~Bolcskei, D.~Gesbert, and A.~J. Paulraj,
\newblock ``On the capacity of {OFDM}-based spatial multiplexing systems,''
\newblock {\em {IEEE} Trans. Comun.}, pp. 225--234, Feb. 2002.

\bibitem{garg05}
P.~Garg, R.~K. Mallik, and H.~M. Gupta,
\newblock ``Performance analysis of space-time coding with imperfect channel
  estimation,''
\newblock {\em IEEE Trans. Wireless Commun.}, vol. 4, pp. 257--265, Jan. 2005.

\bibitem{Biglieri_jour}
Giorgio Taricco and Ezio Biglieri,
\newblock ``Space-time decoding with imperfect channel estimation,''
\newblock {\em IEEE Transaction on Wireless Communications}, vol. 4, no. 4, pp.
  1874, 1888 2005.

\bibitem{sadough06}
S.~Sadough, P.~Piantanida, and P.~Duhamel,
\newblock ``Achievable outage rates with improved decoding of {BICM}
  {M}ultiband {OFDM} under channel estimation errors,''
\newblock in {\em Asilomar Conf. on Signals, systems and computers}, Oct. 2006.

\bibitem{boutros00}
J.~J. Boutros, F.~Boixadera, and C.~Lamy,
\newblock ``Bit-interleaved coded modulations for multiple-input
  multiple-output channels,''
\newblock in {\em Int. Symp. on Spread Spectrum Techniques and Applications},
  Sept. 2000, pp. 123--126.

\bibitem{bcjr}
L.~Bahl, J.~Cocke, F.~Jelinek, and J.~Raviv,
\newblock ``Optimal decoding of linear codes for minimizing symbol error
  rate,''
\newblock {\em IEEE Trans. on Inf. Theory}, pp. 284--287, March 1974.

\bibitem{norme_mb}
Anuj Batra, Jaiganesh Balakrishnan, and Anad Dabak,
\newblock ``Multiband {OFDM} physical layer proposal for {IEEE} 802.15 task
  group 3a,''
\newblock Tech. {R}ep., IEEE, july 2003.

\bibitem{chanReport}
J.~Foerster,
\newblock ``Channel modeling sub-committee report final,''
\newblock Tech. {R}ep., IEEE802.15-02/490, 2003.

\end{thebibliography}
